\documentclass[prl,twocolumn,superscriptaddress,nobalancelastpage]{revtex4-1}

\usepackage{graphicx}
\usepackage{amssymb}
\usepackage{amssymb}
\usepackage{times}
\usepackage{bbm}
\usepackage{color}

\newcommand{\tr}{\mathrm{tr}}

\newcommand{\nn}[1]{\langle#1\rangle}
\newcommand{\grdeg}{\mathcal{D}}

\begin{document}

\title{Information propagation for interacting particle systems}

\author{Norbert Schuch}
\affiliation{Institute for Quantum Information, California Institute of
Technology, MC 305-16, Pasadena CA 91125, USA}

\author{Sarah K.\ Harrison}
\affiliation{Department of Mathematics, Royal Holloway University of
London, Egham, Surrey, TW20 0EX, UK}

\author{Tobias J.\ Osborne}
\affiliation{Institut f\"ur Theoretische Physik, Leibniz-Universit\"at
Hannover, Appelstr.\ 2, 30167 Hannover, Germany}
\affiliation{Institute for Advanced Study Berlin, 14193 Berlin, Germany}

\author{Jens Eisert}

\affiliation{Institute for Advanced Study Berlin, 14193 Berlin, Germany}
\affiliation{Institute of Physics and Astronomy, University of Potsdam, 14476 Potsdam,
Germany}

\begin{abstract}
We show that excitations of interacting quantum particles in lattice
models always propagate with a finite speed of sound. Our argument is
simple yet general and shows that by focusing on the physically relevant
observables one can generally expect a bounded speed of information
propagation. The argument applies equally to quantum spins, bosons such as
in the Bose-Hubbard model, fermions, anyons, and general mixtures thereof,
on arbitrary lattices of any dimension.  It also pertains to dissipative
dynamics on the lattice, and generalizes to the continuum for quantum
fields.  Our result can be seen as a meaningful analogue of the
Lieb-Robinson bound for strongly correlated models.  
\end{abstract}

\maketitle

How fast can information propagate through a system of interacting
particles? The obvious answer seems: No faster than the speed of light.
While certainly correct, this is not the answer one is usually looking
for.  For instance, in a classical solid, liquid, or gas, perturbations
rather propagate at the speed of sound, which is determined by the way the
particles in the system locally interact with each other, without any
reference to relativistic effects.  We would like to understand whether a
similar ``speed of sound'' exists for interacting quantum systems,
limiting the propagation speed of localized excitations, i.e.,
(quasi-)particles.  For interacting quantum spin systems, such a maximal
velocity, known as the Lieb-Robinson
bound~\cite{LR,Cluster,Review,BurrellEisert}, has indeed been shown. While
it seems appealing that there should always be such a bound, systems of
interacting bosons can show counterintuitive effects, in particular since
the interpretation of excitations in terms of particles is no longer fully
justified; in fact, an example of a \emph{non-relativistic} system where
bosons condense into a dynamical state which steadily accelerates has
recently been constructed~\cite{eisert:2009a}.  This example suggests the
disturbing possibility that our intuition is wrong, and only relativistic
quantum theory can provide a proper speed limit.

There are many important reasons, both theoretical and experimental, to
investigate information propagation bounds in interacting particle
systems. It turns out that such bounds lead directly to important, general
results concerning the clustering of correlations in equilibrium
states \cite{Cluster}.  Lieb-Robinson bounds facilitate the simulatability of
strongly interacting quantum systems---the mere existence of a
Lieb-Robinson bound for a quantum system can be used to develop general,
efficient, numerical procedures to simulate the dynamics of lattice 
models~\cite{Simulation}.  From a more practical perspective, new
experiments allow one to explore the non-equilibrium dynamics of ultracold
strongly correlated quantum particles---bosonic, fermionic, or mixtures
thereof---in optical lattices  with unprecedented control
\cite{Cold,Fermions}. In such experiments, it is important to understand
how the particles move: For example, when studying instances of anomalous
expansion, it is far from clear \emph{a priori} whether it is possible to
identify a meaningful speed of sound at all.

\begin{figure}[t]
 \centering
 \includegraphics[width=0.3\textwidth]{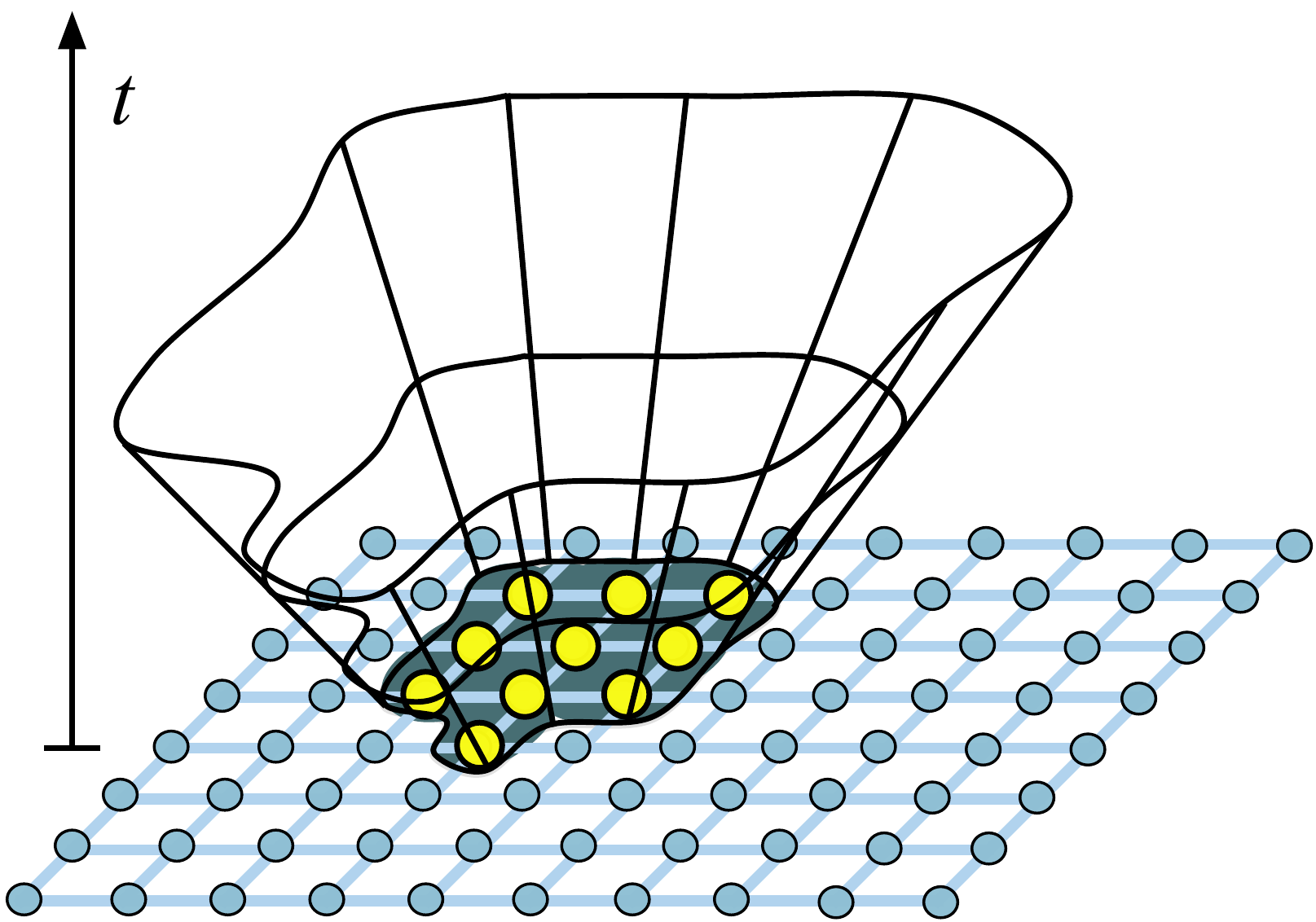}
 \caption{Schematic representation of the ``light cone'' of particles initially placed into a region $R$ 
 of a lattice (yellow circles) and then propagating in time 
 $t$ in a way governed by an interacting quantum model,  outside of which the influence of these particles is exponentially suppressed.}
 \label{fig:projectors}
\end{figure}

The original Lieb-Robinson bound already applies in a very general
setting, namely, to any low-dimensional quantum spin system, and to any
fermionic system confined to a lattice. It is therefore tempting to extend
the original argument to other settings, in particular, to systems of
interacting bosons;  unfortunately, all attempts to do so have run into
insuperable difficulties for systems with nonlinear interactions,
including the Bose-Hubbard model.  The reason for the failure of the
original Lieb-Robinson argument is fundamentally connected to the
unboundedness of the creation operator for bosons: The Lieb-Robinson
velocity depends on the norm of the interaction, which is unbounded for,
e.g., bosons hopping on a lattice, and examples without a speed limit can
be constructed~\cite{eisert:2009a}.

In this Letter, we show how these difficulties can be overcome by
considering the right question concerning the propagation of information.
Our approach allows us to determine Lieb-Robinson type bounds for the
maximal speed at which information can propagate through systems of
interacting particles in a very general scenario: In particular, it
applies to systems of interacting bosons, as well as to fermions, spins,
anyons, or mixtures thereof, both on lattices and in the continuum.
Moreover, it can also be applied beyond Hamiltonian evolution, such as to
systems evolving under some local dissipative dynamics.

The type of system we have in mind is exemplified by the
\emph{Bose-Hubbard model}, a model of bosons hopping on an arbitrary
lattice $G$ of any finite dimension and interacting via an on-site
repulsion,
\begin{equation}\nonumber
	\hat H_{\text{BH}} = -\tau\sum_{\langle j, k\rangle} 
        (\hat b_j^\dag \hat b_k^{\phantom\dag} + \mathrm{h.c.} )
	+ \frac{U}{2}\sum_{j} \hat n_j(\hat n_j-1) - \mu\sum_j \hat n_j,
\end{equation} 
where the first summation is over neighboring sites on the lattice, $\hat
b_j$ is the boson annihilation operator for site $j$, and $\hat n_j =\hat
b_j^\dag \hat b_j$ is the number operator. The natural distance in the
lattice will be denoted by $d(\cdot,\cdot)$, e.g., $d(j,k)=|j-k|$ for a
one-dimensional chain. While we will, for clarity, focus our discussion on
the Bose-Hubbard model, our arguments directly generalize to models of the form
\begin{equation}
\label{eq:generalized-hubbard-model}
    \hat H = -\tau\!\sum_{s=1}^S
    \sum_{ \langle j, k\rangle} 
    (\hat b_{s,j}^\dag \hat b_{s,k}^{\phantom\dag} + \mathrm{h.c.} )
    +f( \{\hat n_{1,j} ,\dots, \hat n_{S,j}\}_{j\in G} )\ ,
\end{equation} 
where the $\hat b_{s,j}$ are annihilation operators for bosons, fermions, or
even anyons of species $s=1,\dots, S$ at site $j$, and $\hat n_{s,j}=\hat
b_{s,j}^\dagger \hat b_{s,j}^{\phantom\dagger}$; the species could for
instance refer to an internal spin degree of freedom.  The interaction
between the particles is characterized by $f$ which can be an arbitrary
function of the local densities, and may involve higher moments of the
particle number, or even non-local interactions. Moreover, our argument
also applies to time-dependent Hamiltonians of this form, as long as the
tunneling amplitude $\tau(\cdot)$ is bounded.

The scenario we consider is described by the Bose-Hubbard model on a
lattice $G$, where in the initial state all sites are empty (i.e., $\langle
\hat n_j\rangle=0$) except for the sites in a region $R$ which can be in an
arbitrary initial state with finite average particle number.
Note that the region $R$ may very well
encompass the major part of the lattice. What we are interested in is how
fast these bosons will travel into the empty part $G\backslash R$ of the
lattice, as a function of the distance $d(\cdot,\cdot)$ on the underlying
graph. In particular, we would like to find a ``speed of sound'' for the
bosons, that is, a velocity $v$ such that for any region $S$ in
$G\backslash R$ with $d(S,R)\ge l$ [i.e.: $d(s,r)\ge l\ \forall s\in
S,\,r\in R$], and for all times $t$ for which $vt<l$, the
expectation value of any observable $\hat O_S$ on $S$ is equal to the
expectation value of the vacuum, up to a correction which decays
exponentially away from the light cone, $e^{\gamma(vt-l)}$.

To start, we consider the Bose-Hubbard model $\hat H_\mathrm{BH}$ and
focus on measurements of the local particle number operators $\hat n_j$.
This corresponds to looking for bosons at the initially empty sites, and
thus captures the most natural notion of particles propagating into a
region.  Let us denote the initial state by $\rho(0)$, which evolves
according to
\[
\dot\rho(t) = -i [\hat H_\mathrm{BH},\rho(t)]
\]
for $t\ge0$. As we are interested in the speed at which particles in the
Bose-Hubbard model propagate, let us try to understand how the local
particle densities
\[
	\alpha_j(t) = \tr(\hat n_j\rho(t)), \quad j\in G\ ,
\]
evolve under $\hat H_\mathrm{BH}$. To this end, we derive a
bound on the rate at which $\alpha_j(\cdot)$ changes, 
which in turn leads to a bound on the velocity at which particles can propagate through
the system. It holds that
\begin{eqnarray}
        \nonumber
	\dot{\alpha}_j(t) &= &
        -i\,\tr\big(\hat n_j\, [\hat H_{\text{BH}},\rho(t)]\big) 
         = 
        -i\,\tr\big([\hat n_j, \hat H_{\text{BH}} ]\,\rho(t)\big) \\
        \label{eq:commutator-bound}
        & =& 2 \tau \sum_{\nn{j,k}}
        \mathrm{Im}\Big[
            \tr\big(\hat b_k^\dag \hat b_j \rho(t)\big)
            \Big]\ ,
\end{eqnarray}
where the summation runs over all sites $k$ neighboring $j$, $d(j,k)=1$.
Since we are only interested in an upper bound on this rate of change, we
now consider $|\dot{\alpha}_j(t)|$ and apply the triangle inequality to
obtain
\begin{equation}
\label{eq:firstbound}
|\dot{\alpha}_j(t)| \le 2\tau\sum_{\nn{j,k}}
        \big|\tr(\hat  b_k^\dag \hat  b_j \rho(t))\big|\ .
\end{equation}
To bound this term we use the operator Cauchy-Schwarz inequality, viewing
\[
	\tr(\hat b_k^\dag \hat b_j \rho(t))= 
        \langle \hat b_k \rho^{1/2}(t), \hat b_j \rho^{1/2}(t) \rangle
\]
as a Hilbert-Schmidt scalar product of $\hat b_j \rho^{1/2}(t)$ and $\hat
b_k \rho^{1/2}(t)$, where $\rho^{1/2}(t)$ is the matrix square root of
$\rho(t)$. This gives rise to
\[
\big|\tr(\hat b_k^\dag \hat b_j \rho(t))\big| \le 
\left({\tr(\hat b_k^\dag \hat b_k^{\phantom\dag} \rho(t))
        \tr(\hat b_j^\dag \hat b_j \rho(t))}\right)^{1/2} .
\]
Combining this with Eq.~(\ref{eq:firstbound}), we
obtain a set of coupled differential inequalities
\begin{equation}
    \label{eq:original-diffineq-system}
    |\dot{\alpha}_j(t)| \le 2\tau\sum_{\nn{j,k}}
	\left({\alpha_j(t)\alpha_k(t)}\right)^{1/2}\ ,
\end{equation}
which, using $\sqrt{xy}\le (x+y)/2$, yields the linearized system
\[
|\dot{\alpha}_j(t)| \le 
    \tau \bigg(
    \grdeg\: \alpha_j(t)+ \sum_{\nn{j,k}} \alpha_k(t)
    \bigg)\ ,
\]
where $\grdeg$ is the maximal vertex degree of the interaction graph.

We are interested in the \emph{worst-case} growth of $\alpha_j(t)$ as $t$
progresses. This will occur when we have equality in the above expression
(i.e., the derivative is as large as possible), and thus a bound
$\gamma_k(t)\ge\alpha_k(t)$ is given by the solution of the linear system
of differential equations 
\[
\dot{\gamma}_j(t) = \tau 
    \bigg(\grdeg\:\gamma_j(t)+	
    \sum_{\nn{j,k}} \gamma_k(t)\bigg)
\]
which fulfills $\gamma_j(0)=\alpha_j(0)$.
This solution has the form
\[
   \vec\gamma(t) = e^{\grdeg\tau t} e^{\tau M t}\,\vec\gamma(0),
\]
where $M$ is the adjacency matrix of the lattice, i.e., $M_{j,k}=1$ if
$d(j,k)=1$ and $0$ otherwise, and $\vec\gamma:=(\gamma_k)_{k\in L}$.
This yields an upper bound
\[
\vec\alpha(t) \le e^{\grdeg\tau t}e^{\tau M t}\,\vec\alpha(0)
\]
for the expected particle number at time $t$ for any site, for $\vec\alpha:=(\alpha_k)_{k\in L}$.

In order to understand how quickly particles propagate from the initially
occupied region $R$ into a region $S$ with $d(R,S)\ge l$, we need to
consider the off-diagonal block of $e^{\grdeg\tau t}e^{\tau Mt}$
corresponding to those two regions.  Thus, in order to obtain a light cone
with an exponential decay $\exp(vt-l)$ outside it, we need to understand
how rapidly the off-diagonal elements of the banded matrix $M$ grow under
exponentiation $e^{\tau Mt}$.
This can be done by applying Theorem 6 from Ref.~\cite{Harmonic}, which
yields for the $(i,j)$-th element of $\exp(\tau M t)$ the bound%
\[
\left[\exp(\tau M t)\right]_{i,j}
    \le	C e^{v_0 t - d(i,j)}
\]
with velocity $v_0=\chi \Delta\tau$, where $\chi\approx
3.59$ is the solution of $\chi\ln\chi=\chi+1$, $\Delta=\|M\|_\infty/2$
depends on the lattice dimension, and $C=2\chi^2/(\chi-1)\approx10$.
Together with the prefactor $\exp(\grdeg\tau t)$, this gives a Lieb-Robinson
velocity $v=v_0+\grdeg\tau$ \cite{NewFootnote}.
For the scenario of an empty lattice with
particles initially placed in a region $R$, this implies that for any $j$ with
$d(j,R)\ge l$,
\begin{equation}
\label{eq:firstmom-lr-bound}
\alpha_j(t) \leq C e^{vt-l} \sum_{k\in R} \alpha_k(0) = C N_0\, e^{vt-l} \ ,
\end{equation}
i.e., up to an exponentially small tail, the particles propagate with a
speed no faster than $v$, independent of their initial state.  Here,
$N_0=\sum_{k\in R}\alpha_k(0)=\langle \hat N\rangle$ is the total number
of particles in the system (i.e., the expectation value of the total
particle number operator $\hat N= \sum_j \hat n_j$). Note that while this (unsurprisingly) means
that the strength of the signal observed may depend on the number of
bosons initially put into the system, the maximum propagation speed $v$
does not depend on $N_0$. In fact, for a purely harmonic one-dimensional 
model for $U=0$, the exact speed of sound is indeed linear in $\tau$, so the above
bound is tight up to a small constant prefactor.

Having understood how to obtain a bound on the propagation speed of
particles, we now turn to more general observables.  First, let us
show how we can bound the higher moments of the particle number operator. For $p\geq 1$,
\begin{eqnarray}
\alpha_j^{(p)}(t) &=&
    \tr\big(\hat n_j^p\rho(t)\big)
    \nonumber
\\
    &=&
    \sum_N\tr\big(\hat n_j \hat n_j^{p-1} P_N \rho(t) P_N\big)
    \nonumber
\\
    &\le &
    \sum_N\tr\big(\hat n_j N^{p-1} P_N \rho(t) P_N\big)
    \label{eq:higher-moments-bnd}
\\
    &\stackrel{(\ref{eq:firstmom-lr-bound})}{\le}&
    \sum_N N^{p-1} \left(C N e^{vt-l}\right) \tr(\rho(t))
    \nonumber
\\
    &=& C\, \langle \hat N^p\rangle\; e^{vt-l}\ ,
    \nonumber
\end{eqnarray}
where $P_N$ projects onto the subspace with a total of $N$ particles, and
we have used that Eq.~(\ref{eq:firstmom-lr-bound}) applies to each
subspace with fixed particle number independently as the Hamiltonian
commutes with $P_N$.  Here, $\langle \hat N^p\rangle$ denotes the
(time-independent) expectation value of the $p$-th moment of the total
particle number operator.  This proves a Lieb-Robinson bound for the
higher moments of the particle number operator.

Let us now turn our attention towards arbitrary local observables $\hat A_j$.
Any such observable can be written as
$\hat A_j=\sum_{p,q} c_{p,q}(\hat b_j^\dag)^p_{\phantom j}\hat b_j^q$, and we
have thus that 
\begin{eqnarray}
\big|\tr(\hat A_j\rho(t))\big| &\le &
    \sum_{p,q} |c_{p,q}|\,\big|\tr[(\hat b_j^\dag)^p_{\phantom j}\hat b_j^q\rho(t)]\big|
\\
    &\le&
    \sum_{p,q} |c_{p,q}|\,\left({\tr\big[(\hat b_j^\dag)^p_{\phantom j} 
		    \hat b_j^p\rho(t)\big]\,
	\tr\big[(\hat b_j^\dag)^q_{\phantom j} \hat b_j^q\rho(t)\big]}\right)^{1/2}\ .\nonumber
\end{eqnarray}
In turn, for $p>0$,
\begin{eqnarray}
    \tr\big[(\hat b_j^\dag)^p_{\phantom j} \hat b_j^p\rho(t)\big]\nonumber
&=&
    \tr\big[\hat n_j(\hat n_j-1)\cdots (\hat n_j-p+1)\rho(t)\big]
\\
&= &\sum_{r=1}^p d_{r,p}\alpha^{(r)}_j(t) \le \tilde C_p e^{vt-l}
\end{eqnarray}
by virtue of Eq.~(\ref{eq:firstmom-lr-bound}), for some constant $\tilde
C_p$. If $p=0$, we trivially have $\tr[\rho(t)]=1$. Together, this yields
a bound 
\[
\big|\tr(\hat A_j\rho(t))\big| \le C' e^{vt-l}
\]
if $c_{0,q}=c_{p,0}=0$ for all $p$ and $q$, and 
\[
\big|\tr(\hat A_j\rho(t))\big| \le C' e^{(vt-l)/2}
\]
otherwise, where we have assumed that $\sum |c_{p,q}|$ is finite, and used
that w.l.o.g.\ $c_{0,0}=0$. In both cases, this means that outside the
light cone given by $vt=l$, $\tr(\hat A_j\rho(t))$ decays exponentially;
however, the decay is on double the length scale in the latter case.

Finally, observables acting on more than one site can be bounded
analogously to the local case: Any two-site operator acting on sites
$j$, $k$ can be written as the sum of terms $\hat A_j \hat A_{k}$, and 
\[
\big|\tr(\hat A_j \hat A_k\rho(t))\big|
\le \left({\tr(\hat A_j^\dagger \hat A_j^{\phantom\dagger}\rho(t))
\tr(\hat A_k^{\phantom\dagger} \hat A_k^\dagger \rho(t))}\right)^{1/2}\ .
\]
The terms on the r.h.s.\ are local observables which can be bounded as
before by $\exp(vt-l)$, yielding the same exponential bound for
two-site---and recursively for many-site---observables. 
(Note that there exist cases where terms which are bounded by
$\exp[(vt-l)/2]$ only appear, and in addition one of the $\hat A$'s above could be
the identity. Thus, bounds of the form $\exp((vt-l)/\kappa)$ can occur,
where $\kappa$ can grow exponentially in the block size. This, however,
still implies that the signal is exponentially small outside the light
cone.)

While we have illustrated our arguments for the Bose-Hubbard model, they
generalize straightforwardly to the more general class of models described
by Eq.\ (\ref{eq:generalized-hubbard-model}). First, it is clear that we can
replace the on-site replusion and chemical potential in the Bose-Hubbard
model by any type of interaction (even a non-local one) which only depends
on the particle numbers, since any such term vanishes in the commutator
$[\hat n_j,\hat H]$ in Eq.~(\ref{eq:commutator-bound}). Second, for
systems that contain several types of bosons the same arguments apply:
Such systems can be modelled using multiple copies of the original graph,
each of which supports the hopping of one individual boson species, and
one obtains independent differential inequalities for the particle
densities $\alpha_{j,s}(t)=\tr[\hat n_{j,s}\rho(t)]$ for each species.

Beyond general bosonic models, our arguments also apply to fermions and
mixtures of bosons and fermions~\cite{BFHubbard}, and in fact even to
anyonic systems.  Again, in a first step one can decouple the individual
species of
particles (which mutually commute) to hop on independent graphs. Then, it
is easy to check that our arguments work independently of the statistics
of the particles, since $[\hat n_j, \hat H]$ in
Eq.~(\ref{eq:commutator-bound}) evaluates to the same expression in terms
of the fermionic (anyonic) creation and annihilation operators.
Even better, fermionic and anyonic systems yield stronger bounds for the
higher moments, and thus for the scenario of general local observables:
In Eq.~(\ref{eq:higher-moments-bnd}), $\hat n_j^{p-1}$ can be bounded by
$1$ instead of $\hat N^{p-1}$, which yields a bound $\alpha_j^{(p)}(t)\le C
N_0\, e^{vt-l}$ on the higher moments.  Corresponding results also follow
for spin systems, as these can be described as hardcore bosons.

Our arguments work not only for unitary theories, but also for certain
types of dissipative (Markovian) models, extending~\cite{Poulin} to
bosonic systems. For instance, in the practically relevant case of a
bosonic system with particle losses, we have that 
\[
\dot \rho(t) = 
    -i\,[\hat H_{\text{BH}},\rho] 
    -\lambda \sum_{j} \left(
    \{\hat b_{j}^\dagger \hat b_{j}^{\phantom\dagger},\rho(t)\}
        -2 \hat b_{j}^{\phantom\dag} \rho(t) \hat b_{j}^\dagger
\right)\ .
\]
Therefore,
\[
    \dot{\alpha}_{j}(t) 
= 
    -i\, \tr([\hat n_{j}, \hat H_\mathrm{BH} ]\rho(t)) 
    -\lambda\,\tr\big(\hat n_{j}\rho(t)\big)\ ,
\]
which shows that the contribution from the dissipative term to
$\dot\alpha_j$ is negative; thus, tighter differential inequalities
and thus a lower speed of sound than in the Hamiltonian case can be
obtained.

To conclude, we have proven that there is a speed limit for the
propagation of information in a system of interacting particles. This
result is particularly relevant for the case of bosons on a lattice, as
bosonic systems cannot be assessed using the established techniques of
Lieb-Robinson bounds due to the unboundedness of the bosonic hopping
operator.  Our argument applies equally to bosonic, fermionic, anyonic,
and spin systems, as well as mixtures thereof, with arbitrary interaction
terms between the particles, and can be generalized to also address
systems with dissipation.  

The key point that allowed us to make statements about the propagation of
information in bosonic systems beyond Lieb-Robinson bounds was first to
focus on a subset of observables relevant to detecting the propagation of
particles, namely the number of particles present at each site, and second
to devise a closed system of inequalities bounding the evolution of their
expectation values.  This allowed us to reduce the problem of
characterizing the full dynamics of the system, which takes place in a
superexponentially large Fock space, to simply keeping track of the dynamics of a
relatively small number of parameters.  This considerably reduced the
complexity of the problem and gave rise to an exactly solvable worst-case
bound.  

The idea of studying information propagation by restricting to a specific
set of observables and investigating the resulting worst-case differential
equation can also be applied to the study of continous systems. This can
be done either by taking an appropriate continuum limit of a lattice
model, or by directly considering a corresponding differential equation
for the particle density which is continuous in space.

{\it Acknowledgements.---}This work was supported by the EU (COMPAS,
MINOS, QESSENCE), the EURYI, the Gordon and Betty Moore Foundation through
Caltech's Center for the Physics of Information, the National Science
Foundation under Grant No.\ \mbox{PHY-0803371}, and the ARO under Grant
No.\ \mbox{W911NF-09-1-0442}.  Part of this work was done at the
Mittag-Leffler-Institute.


\begin{thebibliography}{99}
\bibitem{NewFootnote}
	There are two ways to obtain better bounds on the velocity. First, we can
	bound $\sqrt{xy}\le \frac12(\lambda x+y/\lambda)$, which gives a velocity
	bound $\lambda v_0+\grdeg\tau/\lambda$ for any $\lambda>0$. Second, one can
	solve the non-linear differential
	inequality~(\ref{eq:original-diffineq-system}) by substituting
	$\alpha_j(t)=:\beta_j(t)^2$, which gives linear inequalities
	\[
	\dot{\beta}_j(t) \le \tau \sum_{\nn{j,k}} \beta_k(t)
	\]
	with initial conditions $\beta(0)=\sqrt{\alpha(0)}$.
	The worst-case solution of this system is
	$\vec\beta(t) = e^{\tau M t}\,\vec\beta(0)$,
	which, using the previous estimate of $e^{\tau M t}$, yields a velocity
	$v_0$. In order to obtain bounds on $\alpha_j(t)$, we need to square this
	bound. On the one hand, this implies that the correlations outside the
	light cone decay as $e^{2(vt-l)}$; however, it also yields an
	unfavorable dependence of the prefactor on the initial conditions, 
	$(\sum_j \sqrt{\alpha_j(0)})^2$, which can diverge for a fixed
	number of bosons as the region $R$ grows.
\end{thebibliography}
\end{document}